\documentclass{ws-p8-50x6-00}
\def\today{November 4, 2001}
\begin{document}

\title{HEAVY-FLAVOURED JETS AT HERA}
\author{SANJAY PADHI}
\author{(On behalf of the ZEUS collaboration)}
\address{Department of Physics, McGill University, 
Montreal, \\ Quebec, Canada, H3A 2T8 \\
E-mail: Sanjay.Padhi@desy.de}


\maketitle

\abstracts{ Heavy-flavoured jets have been studied in
  photoproduction at HERA. This includes the first 
   measurement of  dijet angular distributions in D* photoproduction and 
  the ratio of the vector/(vector + pseudoscalar) production rate
  for charm mesons.}

\section{Introduction}
The study of heavy-flavoured jets allows a detailed investigation of pQCD and
the potential to understand the heavy quark production mechanism. 
In this paper only charm is considered. Measurements relating to the
universality of charm fragmentation fraction are made. Measurements of
both inclusive
jets and dijets with a charm meson have been performed to probe
the dynamics and the contents of the photon and proton at HERA.
\vspace {-0.2cm}
\section{Heavy Flavour (charm) Production and Fragmentation}
Charm mesons in the ground state can be in either a vector state (V)
or a pseudoscalar state (PS) corresponding to
D* or D meson\footnote{D$^{*\pm}$(2010) is referred to as D* for the
  rest of this paper.} with
spin 1 or 0, respectively. Their relative production P$_{v}$ = V/(V
+ PS) is sensitive to non-perturbative effects in the hadronisation
process and cannot be calculated exactly. The value P$_{v}$ is
calculated with respect to the ground-state charm mesons via the decay
channels D$^{*+}$
$\rightarrow$ D$^{0}\pi^{+}_{s} \rightarrow (K^{-}\pi^{+})\pi^{+}_{s}(+c.c.)$
and D$^{0}\rightarrow K^{-}\pi^{+}(+c.c.) $. It is assumed that
the D$^{*\pm}$ and D$^{*0}$ are produced with
equal probabilities, all D and D$^{*}$ are produced
only in fragmentation and D$^{0}$ is produced either directly 
(f(c) $\rightarrow$ D$^{0}$ +
X) or via D* decays. This leads to:
\begin{equation}
P_{v} = \frac{1}{(\sigma_{tot}(D^{0})/\sigma_{tot}(D^{*+})) -
  BR(D^{*+} \rightarrow D^{0}\pi^{+} )}
\label{eq:1}
\end{equation}
where BR is the branching ratio and $\sigma_{tot}$ is the measured
cross section. 

 In this paper, the analysis~\cite{1} is based on D$^{*\pm}$ and
 D$^{0}$ events
 with an almost real photon (virtuality, $Q^2_{median}
\approx 3.10^{-4} $ GeV$^2$) in a photon-proton center of mass energy,
W, in
the range $130 <$ W $< 295$ GeV. Using the $\triangle$M = M(D*) -
M(D$^0$) tag the
sample is then further divided into D$^{0}$ mesons arising from and
not from D*
mesons. After this division there were $ 1180 \pm 39$ events with a D$^{0}$
meson from a D* and $ 5223 \pm 185$ inclusive D$^{0}$ meson events and
the resulting value for P$_{v}$ in the full phase space is:
\begin{center}
 P$_{v}$ = 0.546 $\pm 0.045(stat)^{+0.028}_{-0.028}(syst)$.
\end{center}
This is in good agreement with the values of $ 0.57 \pm 0.05 $
and $0.595 \pm  0.045 $~\cite{2} measured in e$^+$e$^-$ annihilations.
\section {Charm with Jets}
In order to understand charm production further, the inclusive jet
cross section for events containing a D* meson has been measured. 
Here the non-perturbative effects, which are currently poorly
understood, are more suppressed than in inclusive charm production. 
Cross sections as a function of the pseudorapidity ($\eta^{jet}$) for
\vspace{-0.2cm}
\begin{figure}[h]
\psfig{figure=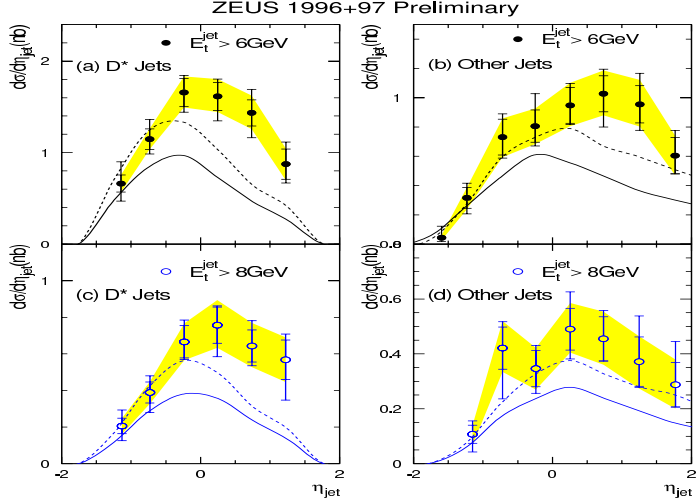,height=2.6in,width=3.6in}
\vspace{-0.3cm}
\caption {\it{  D* and non-D* jet cross sections $d\sigma/d\eta_{jet}$ of D*
  photoproduction events with the kinematic cuts: $Q^2 < 1 $ GeV$^{2}$, 
  $130 < W < 280 $ GeV, p$^{D*}_{\bot} > 3$ GeV and $|\eta^{D*}| < 1.5$. The
  upper plots (a) and (b) are for $E^{jet}_{t} > 6$ GeV and the lower
  plots (c) and (d) are for  $E^{jet}_{t} > 8 $ GeV. The inner error
  bars represent the statistical errors, and the outer bars the
  quadratic sum of statistical and systematic uncertainties. The solid and
  dotted curves are NLO predictions using renormalisation and  
  factorisation scales set to $\mu_{R} = m_{\perp} ; \mu_{F} 
  = 2m_{\perp}$ with charm mass $m_{c} = 1.5$ GeV and $\mu_{R} =
  0.5m_{\perp}$, $ m_{c} = 1.2$ GeV respectively. \label{fig:1}}}
\end{figure}
\vspace{-0.14cm}
D* jet and non-D* jet samples are shown in Fig.~1. The D* jet is
defined to be that jet nearest to the D* in $\eta - \phi
$ ($\triangle$R $\equiv \sqrt{(\phi_{jet} - \phi_{D*})^{2} + (\eta_{jet} -
\eta_{D*})^{2}} < 0.6 $) space. The remaining jets in the event are
called non-D* jets. In Fig.~1, the NLO (``massive
charm'' scheme~\cite{3}) calculations underestimate both
jet cross sections by
approximately a factor of two. Even with an extreme set of
parameters, NLO fails to describe both the shape and the absolute
cross section. The differences in shape and normalization
for both D* and non-D* jet cross sections
cannot be accounted for by the hadronisation corrections.
\vspace {-0.18cm}
\section {Charm with Dijets}
Given the discrepancies between data and the NLO prediction, it is of
interest to probe the kinematics of
charm production in more detail, by measuring the dijet cross
section. The variable ($x^{OBS}_{\gamma}$) related to the momentum
fraction of the parton from the photon, is defined as 
the fraction of the photon's energy participating in the production of
the two highest transverse energy jets:
\vspace {-0.2cm}
\begin{equation}
x^{OBS}_{\gamma} = \frac{\sum_{jet1,2} E^{jet}_{T} e^{-\eta^{jet}}}{2yE_{e}}
\label{eq:2}
\end{equation}
where $yE_e$ is the initial photon energy. 
The measured differential cross section, $d\sigma/dx^{OBS}_{\gamma}$ is
compared with HERWIG MC (normalised to the data) and predictions from
NLO massive charm calculation~\cite{3} in Fig.~2.
There is a  substantial tail at low $x^{OBS}_{\gamma}$, which 
requires a LO-resolved component to describe the
data. The predictions from NLO, where
charm is not treated as an active flavour in the photon structure
function, significantly underestimates the data at low
$x^{OBS}_{\gamma}$. Using an extreme set of parameters 
in the calculation yields a larger cross section at low
$x^{OBS}_{\gamma}$, but is still below the data. 
 Further studies were made to probe more directly the production mechanism.
 This was done by considering the differential distribution~\cite{4} $
 dN/d|\cos\theta^{*}|$, where $\theta^{*}$ is the angle between the
 jet-jet axis
 and the beam direction in the dijet rest frame. This distribution is
 sensitive to the parton dynamics of the underlying sub-processes. The
 distribution was considered for direct-enriched ($x^{OBS}_{\gamma} >
 0.75$) and resolved-enriched ($x^{OBS}_{\gamma} < 0.75$) samples.
 Additional cuts (as shown in Fig.~3) on the dijet invariant mass,
 M$_{jj}$, and 
 the average pseudorapidity of the jets, $|\overline{\eta}|$, were
 checked to ensure an unbiased phase space region.
 The measured differential distributions $dN/d|\cos\theta^{*}|$ for
  both $x^{OBS}_{\gamma} < 0.75 $ and $x^{OBS}_{\gamma} > 0.75 $ are
  shown in
 Fig.~3. They are significantly different, which reflects
 the different spins of the quark and gluon propagator. This is
 well reproduced by the PYTHIA prediction.
 The steep rise towards high $|\cos\theta^{*}|$ values of the resolved
 dijet charm events, consistent with gluon exchange, provides strong
 evidence that the bulk of the resolved contribution is
 due to charm excitation in the photon, rather than to the more
 conventional resolved process $gg \rightarrow c\overline{c}$.
\begin{figure}[h]
\unitlength1cm  \begin{picture}(2,8)
\includegraphics{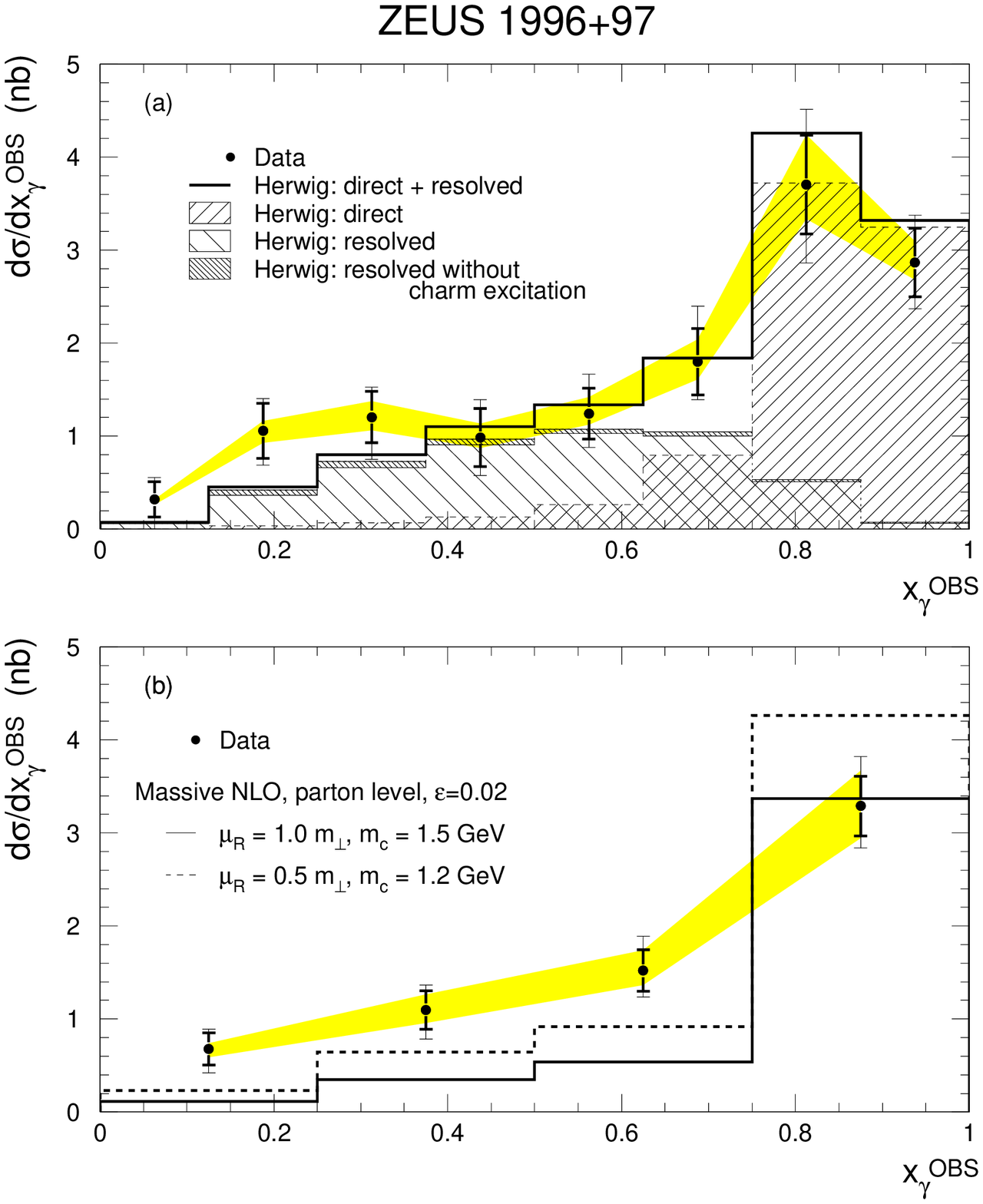}
\includegraphics{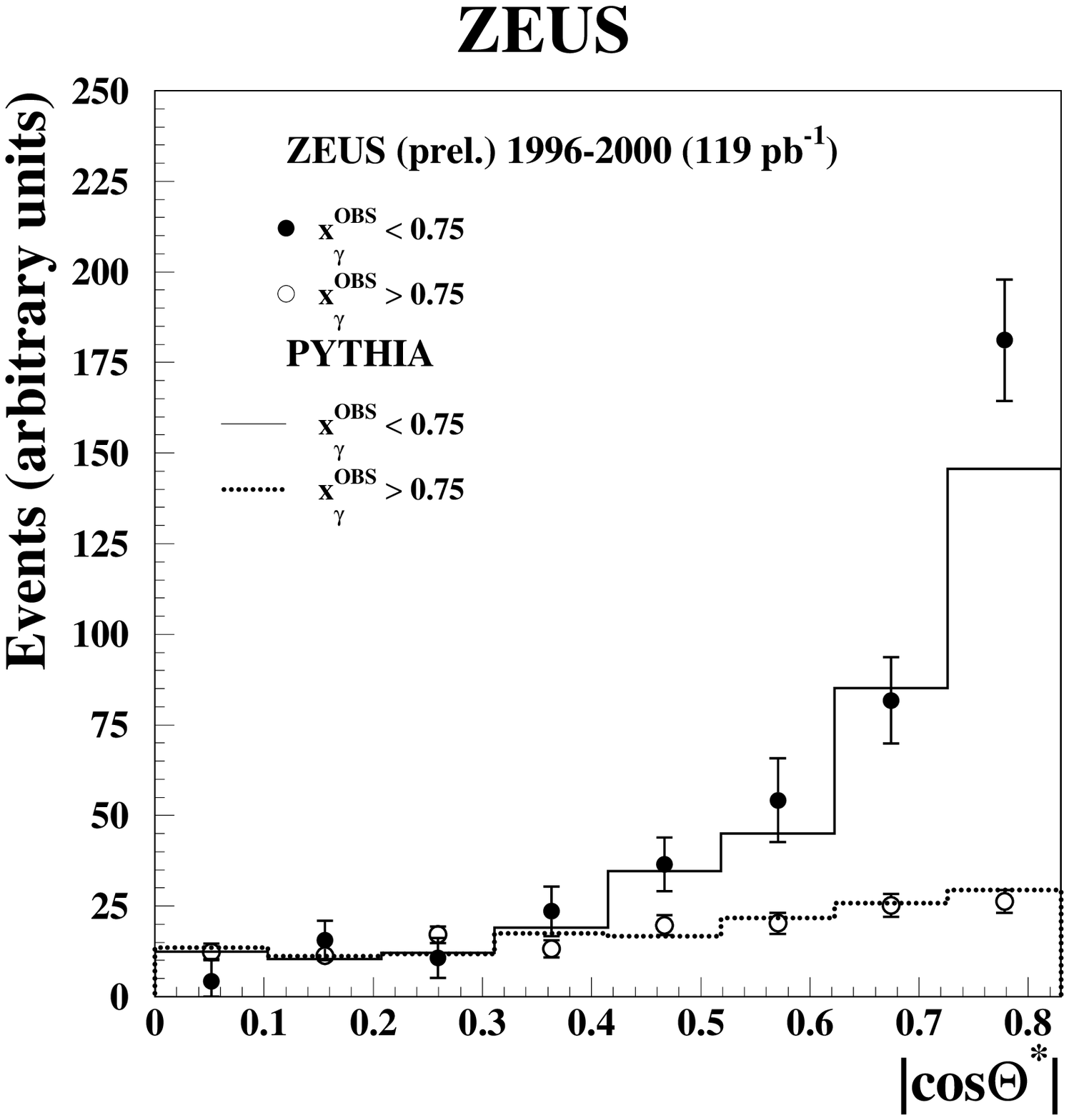}
\end{picture}
\vspace{-2.7cm}
\caption{\it (left) The differential cross section
    $d\sigma/dx^{OBS}_{\gamma}$ for dijets with an associated D* meson
    with $p^{D*}_{\bot} > 3$ GeV, $|\eta^{D*}| < 1.5$, 130 $< W <$ 280
    GeV, $|\eta^{jet}| < 2.4$, E$^{jet1}_T > 7$ GeV and E$^{jet2}_T >
    6$ GeV. The experimental data (dots) are compared to (a) HERWIG
    MC and (b) a parton level NLO 
    calculation with the parameters shown in the plot.}
\vspace{-0.3cm}
\caption{\it (right) The differential distributions $dN/d|cos\theta^{*}|$ of
  the data (dots)
    with $p^{D*}_{\bot} > 3$ GeV, $|\eta^{D*}| < 1.5$, 
  130 $< W <$ 280 GeV, $|\eta^{jet}| < 2.4$, E$^{jet1,2}_T > 5$ GeV,
  M$_{jj} >$ 18 GeV, $|\overline{\eta}| <$ 1.2 and
  of PYTHIA MC simulations (lines). Results are given separately for
  direct photon (open dots/dashed lines) and for resolved photon
  (black dots/full histogram) events. All the distributions are
  normalised to the resolved data distribution in the lowest 4 bins.}
\vspace{-0.7cm}
\end{figure}
\section {Conclusions}
\vspace {-0.2cm}
Heavy-flavoured jets provide an important tool for
understanding the heavy quark production mechanism. Measurement
of P$_{v}$
confirms the universality of the charm fragmentation. However
inadequacies in current NLO calculations in describing heavy-flavour
jet production are clearly evident. The
$|\cos\theta^{*}|$ distribution for dijet events with a D* shows a
clear signature of gluon propagator for events with $x^{OBS}_{\gamma}
< 0.75$, suggesting strong evidence of charm in the photon.

\end{document}